\documentclass[12pt]{iopart}
\usepackage{iopams} 
\usepackage[usenames,dvipsnames]{xcolor} 
\usepackage[normalem]{ulem}

\begin{document}

\title[Cosmic censorship and the third law of black hole dynamics]{Cosmic censorship and the third law of black hole dynamics}

\author{Koray D\"{u}zta\c{s}}

\address{Physics Department, Eastern Mediterranean  University, Famagusta, Northern Cyprus, Mersin 10, Turkey}
\ead{koray.duztas@emu.edu.tr}
\vspace{10pt}

\begin{abstract}
Israel's proof of the third law of black hole dynamics is based on the semi-rigidity assumption which requires that the areas of the outermost trapped surfaces are preserved. We argue that the condition that the area of the event horizon should not decrease is necessary -but not sufficient- for the semi-rigidity assumption to hold. It is known that the presence of a naked singularity in the exterior region invalidates Hawking's proof of the area theorem. In this sense, the third law assumes the validity of the strong form of the cosmic censorship conjecture, contrary to general acceptance. Still, one cannot guarantee that the semi-rigidity assumption holds. Therefore the current proof is incomplete. Recent studies on the interactions of black holes with fermionic fields imply a generic formation of naked singularities. The presence of these singularities invalidates the laws of black hole dynamics and cosmic censorship conjecture for the processes that satisfy the weak energy condition.
\end{abstract}

%
\vspace{2pc}
\section{Introduction}
According to the singularity theorems of Hawking and Penrose, the formation of singularities is inevitable as a  result of gravitational collapse, given some very reasonable assumptions~\cite{singtheo}. Whether these singularities can be causally related to distant observers, is one of the fundamental problems of classical general relativity. Cosmic censorship conjecture (CCC) was proposed to disable any contact of these singularities with distant observers~\cite{ccc}. In its  weak form (wCCC) the conjecture asserts that the singularities that arise in gravitational collapse are always hidden behind event horizons. In that case, a distant observer does not encounter singularities or any effects propagating out of them, and the consistency and the deterministic nature of general relativity is reassured at least in the space-time excluding the black hole region bounded by the event horizon.

It has not been possible to establish a concrete proof of CCC. For that reason Wald constructed an alternative thought procedure to test the stability of event horizons~\cite{wald74}. Consider a Kerr-Newman black hole defined by three parameters (Mass $M$, charge $Q$, and angular momentum per unit mass $a$), which should satisfy
\begin{equation}
M^{2} \geq Q^{2}+a^{2}. \label{criterion}
\end{equation}
The condition (\ref{criterion}) ensures that an event horizon exists. Then the black hole is allowed to absorb test particles or fields incident from infinity. The space-time is expected to settle to a new Kerr solution with  parameters $M,Q,a$ at the end of the interaction . If the final configuration of parameters satisfies (\ref{criterion}),  the horizon is stable, and the wCCC remains valid. If the final configuration fails to satisfy (\ref{criterion}), the black hole has turned into a naked singularity and wCCC is violated. 

In the first example of these thought experiments Wald showed that particles with enough charge or angular momentum to destroy the horizon either miss or are repelled by the black hole. Following Wald, many similar thought experiments were constructed to test the validity of CCC in the interaction of black holes with test particles or fields. These problems involve justifications and possible violations of  wCCC, as well as employments of back-reaction effects to restore the horizon~\cite{his,hu,dkn,ma,h2,ri1,ma1,ri2,h1,jac,b1,ov,emccc,su,ha}. 

The most generic violation of CCC occurs in the interaction of black holes with fermionic  fields~\cite{du,to2,nat}. Since fermionic fields do not exhibit superradiant behaviour, the absorption of modes carrying low energies and relatively high angular momentum is allowed. As the frequency $\omega$ of the incoming field is reduced below the superradiant limit, the contribution of the interaction to the  angular momentum of the black hole increases without bound with respect to its contribution to the energy of the black hole. The back reaction effects become negligible as $\omega$ is lowered, and the formation of a naked singularity cannot be avoided. For that reason, the violation of wCCC by fermionic fields is generic as opposed to cases involving test bodies and scalar fields. This should not be confused with the attempts to overspin black holes by quantum tunnelling of a single fermion~\cite{ri1,ma1,ri2}. Spontaneous emission of black holes acts as a cosmic censor to dominate the effect of  a single (or a few) particle by many orders of magnitude, though its effect is negligible against challenging fields~\cite{su,ha}. (see \cite{stab} for a general discussion) The generic violation of wCCC due to the absence of superradiance also applies to the asymptotically anti-de Sitter case~\cite{btz}, and Kerr-Taub NUT black holes\cite{ktn}.  

The third law of black hole dynamics does not allow the black hole to reach extremality if the relevant energy momentum tensor satisfies the weak energy condition. The general acceptance is that the proof of this law does not pre-assume CCC. For that reason it is believed that backreactions effects can be employed to restore CCC, if a violation is observed in a problem involving bosonic matters or fields. The fact that a generic violation occurs in the case of fermionic fields does not contradict general assumptions since the energy momentum tensor does not satisfy the weak energy condition. 

In  this work we envisage that there exists a naked singularity in the exterior region which implies that the strong form of CCC is violated. Restricting ourselves to processes that satisfy the weak energy condition,  we revisit the proofs of Hawking's area theorem and the third law of black hole dynamics to test whether a black hole can be driven to extremality and beyond. We focus on the assumptions leading to Israel's proof of the third law  to examine whether they can be sustained if there exists a naked singularity in the exterior region. It turns out that the third law  as well as Hawking's area theorem assume the validity of the strong form of CCC. Based on these results, we re-evaluate  the validity of CCC in classical general relativity.

\section{Cosmic censorship and black hole dynamics}
The black hole region of a space-time $M$ is defined as the set of all points that cannot be reached by past directed causal curves from future null infinity $\mathcal{I}^+$.

\begin{equation}
\mathcal{B}= M- J(\mathcal{I}^+)
\end{equation}
The event horizon  constitutes its boundary in $M$.
\begin{equation}
\partial\mathcal{B}= \dot{J}^-(\mathcal{I}^+)\cap M
\end{equation}
where $J^-$ denotes the causal past of a set and $\dot{J}^-$ denotes its boundary. By definition, the event horizon is the boundary of the set of all points that can be reached from  $\mathcal{I}^+$ by past directed causal curves. According to the no-hair theorem the metric outside a collapsed object settles down to one of the Kerr family of solution with $M \geq a$. Then the event horizon is topologically a two-sphere with area
\begin{equation}
A=8\pi M\left(M + \sqrt{M^2- a^2}\right)
\label{area}
\end{equation}
Hawking proved that if two black holes capture each other to form a single one, the area of the event horizon of the final black hole is larger than the sum of initial areas [25]. The result also applies to the case of a single black hole interacting with particles or fields. The proof of the area theorem assumes that there are no naked singularities in the exterior region, also the solution in the final case resembles the exterior Kerr solution with no naked singularities, i.e. $M \geq a$ . Naively one expects that the area of the event horizon should not decrease if the relevant process satisfies the weak energy condition. However if there exists a naked singularity in the exterior region, Hawking's proof becomes inconclusive. In principle, the area of the event horizon may decrease whether or not the weak energy condition is satisfied.  In this work we evaluate the case in which there exists a naked singularity in the exterior region. To avoid any confusion we note that this naked singularity is placed sufficiently far from the black hole, so that it can not interact with the black hole to modify or perturb the space-time geometry. The main function of the singularity is to disable the definition of a Cauchy surface for the space-time so that the proof of the area theorem becomes invalid.

The third law was first suggested by Bardeen, Carter and Hawking who were inspired from the analogy between the laws of black hole dynamics and thermodynamics \cite{thermo}. Israel established the formal statement of the law and gave a proof \cite{israel1}. According to the formal statement, a black hole cannot become extremal by losing all of its trapped surfaces in a finite advanced time if the energy-momentum tensor of the perturbation satisfies the weak energy condition in a neighbourhood of the apparent horizon. Israel's proof directly follows from his gravitational confinement theorem which states that a trapped two surface  $S_0$  can be extended into the future, to a three cylinder $\Sigma$  that is everywhere space-like \cite{israel0}. The proof of this theorem is mainly based on the assumption that the extension is semi-rigid; i.e. the evolution of the trapped surface $S_0$  is locally area preserving. In addition, he assumes that $\Sigma$  is regular, it does not encounter any singularities in its development and the weak energy condition holds on $\Sigma$.  There is a subtle difference between Israel’s assumption that the trapped surfaces do not encounter any singularities in their developments and the assumption of cosmic censorship. In this work we envisage that there exists a naked singularity in the exterior region. However, the trapped surfaces do not encounter this singularity in their developments. In other words, though the strong form of CCC is violated, Israel’s assumption of no singularities is preserved.  
 
Though the formal statement of the third law only refers to the assumption of the weak energy condition, the main assumption that leads to the proof of the confinement theorem and the third law is semi-rigidity. To be more precise, consider that $S_0$ is Lie transported along the integral curves of a vector field $\xi^{\alpha}$  parametrized by $t$ with $t=0$ on $S_0$. The variation along the normal direction $n^{\alpha}$ of the element of two-area $dS$ is given by
\begin{equation}
\mathcal{L}_{n^{\alpha}} dS = -\xi^A K_A dS
\end{equation}
where $K_A$ $(A=(0,1))$ are  the traces of the extrinsic curvatures of $S(t)$. Then the semi-rigidity assumption can be stated as 
\begin{equation}
\xi^A K_A =0
\label{rigid}
\end{equation}
One notes that , initially $K_0>0$ and $K_1>0$ always. Then (\ref{rigid}) implies that $\xi^1$ is initially positive, so  $N^{\alpha}N_{\alpha} \equiv -\xi^A \xi_A$ is  initially negative. Israel proceeds to prove that it cannot become positive by passing through zero (See \cite{israel0} for details). Then, the semi-rigid future extension of the trapped surface $S_0$ must remain space-like.

The same proof would also apply to the case where one assumes  $\xi^A K_A <0$ so that the area of $S_0$ is allowed to increase. In that case, $\xi^1$ is still initially positive, and  $N^{\alpha}N_{\alpha} $ is  initially negative. The rest of the proof is identical. Therefore the future extension of the trapped surface $S_0$  remains space-like if $\xi^A K_A \leq 0$; i.e. if the area is preserved or allowed to increase. The proof becomes invalid only if the area decreases $(\xi^A K_A >0)$.  

The  assumption that the area of a trapped surface does not decrease, leads to the proof of the gravitational confinement theorem which states that  an initially trapped surface remains trapped, and the interior of that surface cannot causally influence the exterior world \cite{israel0}. The third law of black hole dynamics is an immediate consequence of this theorem. Consider the foliation of the three cylinder $\Sigma$ by  two sections $S(\tau)$, where $\tau$ is a non-unique time function. The confinement theorem implies that all subsequent two sections $S(\tau)$ are trapped.  If we choose $S_0$ as one of the outermost trapped surfaces and semi-rigidly extend it to the future, the extension should remain trapped. (See \cite{israel1} for details) 

The proof assumes the validity of an area theorem for the outermost trapped surfaces and the apparent horizon as their limit, analogous to Hawking's area theorem for the event horizon. Hawking's area theorem is known to pre-assume the strong form of CCC. A natural question at this stage is whether an analogous area theorem for the apparent horizon can be proved without assuming the validity of CCC.

In an asymptotically predictable space-time any trapped surface $S$ is entirely contained within the black hole region; $S \subset \mathcal{B}$. The apparent horizon of a black hole is defined as the boundary of trapped surfaces. Since the black hole region is bounded by the event horizon, the apparent horizon is necessarily inside the event horizon. At late times, as the black hole settles down, the event horizon and the apparent horizon asymptotically approach each other; they will not be widely separated. Then the areas are related by
\begin{equation}
A_{\rm{EH}}\gtrsim A_{S_0}
\end{equation}
where  $S_0$ is one of the outermost trapped surfaces. The equality is asymptotically approached at late times. The proof of the third law only applies to the cases where the semi-rigidity assumption holds; i.e. the area of the outermost trapped surfaces are preserved, or allowed to increase. For a black hole perturbed by smooth processes at late times, this cannot be achieved if the area of the event horizon  decreases. We pointed out that, in a process satisfying the weak energy condition, the area of the event horizon can decrease  if there exists a naked singularity in the exterior region. To avoid any confusion, we do not assume that the trapped surface encounters this singularity in its development. A singularity  located in the exterior region is sufficient to invalidate Hawking's area theorem. If the area of the event horizon decreases, it will not be possible to extend the outermost trapped surfaces into the future in an area preserving fashion, while they are constrained to remain inside the event horizon. Then, the semi-rigidity assumption cannot be sustained and the gravitational confinement theorem and the third law becomes invalid.  Therefore the third law of black hole dynamics and the gravitational confinement theorem which it relies on, pre-assume the validity of the strong form of CCC.  

Though the strong form of CCC is necessary for the semi-rigidity assumption to hold, one cannot guarantee that it will be sufficient. The fact that the area of the event horizon cannot decrease, does not necessarily imply that the area of the trapped surfaces inside the horizon are also preserved. This has to be proved separately, without making further assumptions.  In this sense the current proof of the third law is incomplete, and it cannot be considered as a rigorous proof.

\section{Conclusions}
The status of the third law of black hole dynamics may be regarded as controversial unlike the other laws which are based on rigorous proofs. There exists works supporting and questioning its validity \cite{dadhich,racz} . In this work we envisaged that there exists a naked singularity in the exterior region and evaluated the processes which satisfy the weak energy condition. As is well known, Hawking's area theorem pre-assumes that no naked singularities exist in the exterior region. We showed that the third law of black hole dynamics also becomes invalid if there exists a naked singularity and the area of the event horizon decreases. Moreover, the semi-rigidity assumption in Israel's proof does not necessarily hold even if the strong form of CCC is valid and the area of the event horizon does not decrease. We conclude that the current proof of the third law assumes the strong form of CCC contrary to general acceptance; and even with this assumption it is incomplete and it cannot be considered as a rigorous proof.

We  mentioned that the interactions of black holes with fermionic fields lead to a generic violation of wCCC to form naked singularities. These interactions may be excluded from the cosmic censorship hypothesis since the energy-momentum tensor does not satisfy the weak energy condition. However,  we have proved that, if such naked singularities can evolve, their presence in the exterior region invalidates not only Hawking's area theorem but also the third law of black hole dynamics. Then it is possible for the processes that satisfy the weak energy condition to drive a black hole to extremality. By analytical continuation of the same process, a black hole can turn into a naked singularity. To summarise, fermionic processes lead to the formation of naked singularities, and the presence of these naked singularities enables the formation of other naked singularities in the processes satisfying the weak energy condition. In this case the weak cosmic censorship conjecture is also violated.

The results in this work do not imply that every process will lead to the destruction of the event horizon if there exists a naked singularity in the exterior region. Rather than that, we proved that one cannot rely on the laws of black hole dynamics or CCC to conclude in advance, that the horizon will be preserved. Even if there exists no naked singularities in the exterior region, only Hawking's area theorem is valid, no rigorous proof exists for the third law or CCC. Still, one can evaluate each process independently to check whether or not the event horizon will be preserved.

\end{document}